# Single atom enables extraordinary light transmission through zero-mode waveguide

V. V. Klimov[1,2] *

[1]Optics department, P.N. Lebedev Physical Institute, Russian Academy of Sciences, 53 Leninsky Prospect, Moscow 119991, Russia

[2]Laboratory of optical methods for DNA molecules sequencing, Institute of Spectroscopy, Russian Academy of

Sciences, Moscow, Troitsk 108840, Russia

*klimov256@gmail.com

Abstract: The integration of elementary quantum objects (atoms, molecules, and quantum dots) with solid-state nanostructures lies at the forefront of nano-optics, nanophotonics, and quantum information science. However, the advancement of this field is hindered by the lack of a rigorous and feasible theoretical framework for describing atom–nanostructure interactions, which are inherently complex and multi-parametric. In this work, we develop a theory of light transmission through a Zero-Mode Waveguide (ZMW) containing a single atom. It is shown that the presence of a single atom inside the ZMW can lead to either a significant enhancement or suppression of light transmission, depending on the detuning of the excitation field frequency from the atomic resonance. This extraordinary transmission and blocking effect can be employed for studying the spatiotemporal dynamics of atoms in complex nanoscopic environments, probing quantum optical phenomena, and developing novel nano-optical devices.

Modern nanotechnology has led to the development of nano-optics, where optical fields can be localized at the nanoscale, that is, at scales much smaller than the wavelength of light in vacuum. In nano-optics, the effects associated with the localization of light in single nanoholes [1] - [3] and the influence of nanoholes on the fluorescence of molecules [4]-[6] are especially interesting. Nanoholes in metal nanofilms can have volumes orders of magnitude smaller than those achievable by confocal microscopy. Because of this, optical radiation cannot propagate in such nanoholes and therefore now they are often called Zero-Mode Waveguides (ZMW). ZMWs open up completely new perspectives for analyzing of the dynamics of molecules based on studying of their fluorescence fluctuations [7]-[18].

However, not only the fluorescence of molecules in ZMW is of significant theoretical and practical interest. The very phenomenon of light transmission through ZMW is also a very complex and interesting phenomenon. The classical theory of light transmission through a single hole in an infinitely thin metal film [1]-[3] shows that the coefficient of light transmission through a nanohole is extremely small.

However, the effect of extraordinary light transmission (EOT) through ZMWs was recently discovered by excitation of surface plasmon resonance. Surface plasmon resonance was excited by a sub-wavelength hole array [19]-[20] or a bull's eye nanostructure around a hole [21]. As a result, the transmission of light increases significantly, and the effective cross section of the individual aperture becomes significantly larger than its physical size. EOT is also possible by combining ZMW and a photonic crystal [22]-[25]. In all these cases, EOT is achieved by focusing the external field by a material structure whose dimensions are large compared to the holes.

The original idea of EOT, or more precisely, transfer, of light through a nanohole using an excited atom, which emits light after passing through a ZMW, was proposed in [26].

In this work, we show that extraordinary light transmission through a nanohole in a free-hanging metal film in a vacuum can be achieved by placing a neutral, unexcited atom in the ZMW without using any additional nanostructures. This statement of the problem is new, since the fluorescence of molecules in ZMW is usually considered, in which there is no interference between the exciting field and the field emitted by the molecule.

The scattering of light by an atom already in free space has a number of remarkable properties. For example, the cross-section of resonant scattering of light by a free atom $\sigma_{res}$ is determined by the resonant wavelength $\lambda_0$ rather than by the Bohr radius $r_{Bohr}$ [27] :

$$\sigma_{res} = \frac{3}{2\pi} \lambda_0^2 \gg r_{Bohr}^2 \qquad (1)$$

It will be demonstrated below that the phenomenon of light transmission through a ZMW with a single atom inside is very complicated and can lead to both a significant blocking and a significant increase in light transmission, that is, to the effect of EOT. The geometry of the problem is shown in Fig. 1.

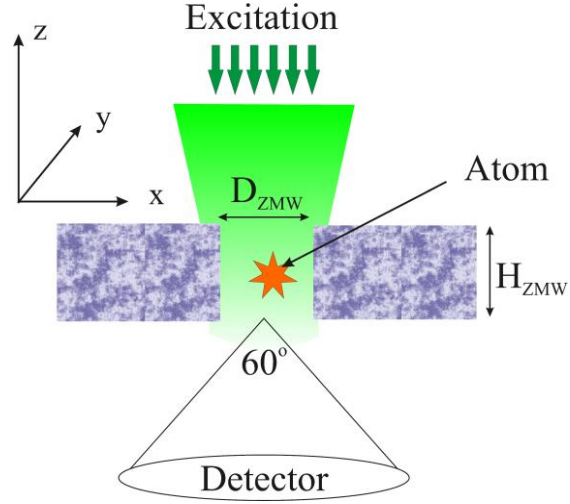

Fig.1. Geometry of the problem. Circularly polarized light impinges normally on ZMW with diameter $D_{ZMW}$ in Al film of thickness $H_{ZMW}$. Transmitted light is collected by a lens with aperture angle of 60 degrees.

First of all, to describe the interaction of an atom with light in ZMW, knowledge of the quantum dynamic dipole polarizability of an atom in a nanoenvironment is required. The dependence of the dynamic dipole polarizability of any quantum object, for example, an atom, $\alpha_{0,A}$, on frequency $\omega$ is characterized by a very general expression (see, for example, [27]):

$$\alpha_{0,A} = \frac{3}{4k_0^3} \frac{\Gamma_0}{(\omega - \omega_0 + i\Gamma_0/2)}, \quad \Gamma_0 = \frac{|\mathbf{d}_{mn}|^2 k_0^3}{3\hbar} \qquad (2).$$

In (2) and below $k_0 = \omega/c$, $\mathbf{d}_{nm}$ denotes the dipole moment matrix element, and $\omega_0$ denotes the resonant frequency. At resonance, (1) can be obtained from (2).

When an atom is in ZMW at the point $\mathbf{r}_A$, the influence of ZMW on the dynamics of atomic electrons must be taken into account, which leads to changes in the linewidth and transition frequency (Purcell effect [5],[28]-[30]). These changes are described by the following expressions

$$\omega_0 \to \tilde{\omega}_0 = \omega_0 - \frac{3\Gamma_0}{4k_0^3} \operatorname{Re} \mathbf{m} \ddot{G}^R_{ZMW}(\mathbf{r}_A,\mathbf{r}_A) \mathbf{m}$$

$$\Gamma_0 \to \tilde{\Gamma}_0 = \Gamma_0 + \frac{3\Gamma_0}{2k_0^3} \operatorname{Im} \mathbf{m} \ddot{G}^R_{ZMW}(\mathbf{r}_A,\mathbf{r}_A) \mathbf{m} \qquad (3)$$

$$\mathbf{m} = \mathbf{d}_{nm}/|\mathbf{d}_{nm}|$$

In (3) $G_{ZMW}^{R}(\mathbf{r},\mathbf{r}')$ denotes the reflected Green's function of the electromagnetic field in ZMW

$$\ddot{G}_{ZMW}^{R}(\mathbf{r},\mathbf{r}') = \ddot{G}(\mathbf{r},\mathbf{r}') - \ddot{G}^{(0)}(\mathbf{r}-\mathbf{r}') \qquad (4)$$

where the full Green's function satisfies the equation:

$$\nabla \times (\nabla \times \ddot{G}(\mathbf{r},\mathbf{r}')) - k_0^2 \varepsilon(\mathbf{r}) \ddot{G}(\mathbf{r},\mathbf{r}') = 4\pi k_0^2 \ddot{1} \delta(\mathbf{r}-\mathbf{r}') \quad (5),$$

and $G_{ij}^{(0)}(\mathbf{r},\omega)$ stands for the Green's function of the free space [31]:

$$\ddot{G}_{ij}^{(0)}(\mathbf{r}) = \left[ k_0^2 \frac{(\delta_{ij} - n_i n_j)}{r} + \frac{(3 n_i n_j - \delta_{ij})}{r^3}(1 - i k_0 r) \right] e^{i k_0 r}, \qquad (6)$$

($\mathbf{n} = \mathbf{r}/r$ is the unit vector in the direction from an atom to the observation point).

It is important that in the presence of the ZMW tensor $\ddot{G}_{ZMW}^{R}(\mathbf{r},\mathbf{r}')$ is not isotropic one and instead of one resonant frequency of an atom, 3 frequencies may appear. However, if the atom is on the symmetry axis of the system, then $\ddot{G}_{ZMW}^{R}(\mathbf{r},\mathbf{r}_0) = diag(G_{ZMW,\perp}^{R}, G_{ZMW,\perp}^{R}, G_{ZMW,\parallel}^{R})$ and the effective polarizability tensor $\ddot{\alpha}_A$

$$\ddot{\alpha}_A = (\alpha_{0,A}^{-1} - \ddot{G})^{-1} \qquad (7)$$

is isotropic tensor in the *x-y* plane, which is perpendicular to the axis of the system.

Knowing the polarizability of an atom in the presence of ZMW, one can find a general expression for the signal in a detector having an area *S*, as the intensity of the total field $\mathbf{E}_{tot}(\mathbf{r}_{det})$, which is the sum of the incident $\mathbf{E}_{0,ZMW}(\mathbf{r}_{det})$ and scattered $\mathbf{E}_{sc}(\mathbf{r}_{det}) = \ddot{G}(\mathbf{r}_{det},\mathbf{r}_A) \ddot{\alpha}_A \mathbf{E}_{0,ZMW}(\mathbf{r}_A)$ fields. In the case of an on-axis atom, all quantities are isotropic in the *x-y* plane and for power $P_{det}$ in a detector of area *S* one can obtain an expression

$$P_{det} \approx P_0 \left| 1 + \xi \alpha_A \right|^2; P_0 = \frac{c}{8\pi} E_{0,ZMW}^2(\mathbf{r}_{det}) S;$$
$$\xi = G_\perp(\mathbf{r}_{det},\mathbf{r}_A) E_{0,ZMW}(\mathbf{r}_A) / E_{0,ZMW}(\mathbf{r}_{det}) \qquad (8)$$

In the radiation zone $\mathbf{r}_{det} \to \infty$ and parameter $\xi$ is determined only by the geometry of the ZMW and its material, and from dimensions considerations one can write $\xi = \bar{\xi} / R_{ZMW}^3$, where $\bar{\xi} \sim 1$. Therefore, to describe the effect of EOT, one can use the formula

$$P_{det}/P_0 \approx \left|1+\bar{\xi}\frac{3}{16\pi^3}\left(\frac{\lambda_0}{R_{ZMW}}\right)^3\frac{\tilde{\Gamma}/2}{(\omega-\tilde{\omega}_0+i\tilde{\Gamma}/2)}\right|^2 \quad (9)$$

Eq.(9) describes the classical Fano resonance [34] and gives a very good quantitative description of the effect of EOT due to the presence of a single atom in the ZMW. In particular, when $\lambda_0 = 532\text{nm}, R_{ZMW} = 50\text{nm}$ from (9) with $\bar{\xi}=1$ one can obtain a qualitative estimate of the maximum transmission coefficient, $P_{det}/P_0 \sim 50$.

In the case of a deep ZMW ($H_{ZMW}\to\infty$) with perfectly conducting walls, instead of (9) one can obtain the exact analytical expression. A general expression for the Green's function in a cylindrical waveguide with perfectly conducting walls can be found, for example, in [35] (Ch.5, eqs. (89)-(90)). In the case where the waveguide depth is much greater than the propagation length of any exponentially decaying mode and when the atom is located near the center of the waveguide, to describe the effect of extraordinary transmission it is sufficient to consider only the mode with the least attenuation. In the case of excitation of the waveguide by a normally incident circularly polarized wave, the TE$_{11}$ decaying mode inside the waveguide will have the least attenuation. By keeping only this mode in the mode expansion of the Green's function ([35],Ch.5, eqs. (89)-(90)) we can obtain an analytical expression for the extraordinary transmission coefficient $P_{det}/P_0$

$$P_{det}/P_0 \approx \left|1+\frac{3z^{*2}}{2k_0 R_{ZMW}(z^{*2}-1)J_1^2(z^*)^2\sqrt{z^{*2}-k_0^2 R_{ZMW}^2}}\frac{\tilde{\Gamma}/2}{(\omega-\tilde{\omega}_0+i\tilde{\Gamma}/2)}\right|^2 \quad (10)$$

where $J_1'(z^*)=0$. From expression (10) it is evident that even in this idealized case the nature of the interaction between the atom and the ZMW has non-trivial dependencies on the geometry of the ZMW and the characteristics of the atom. An important feature of expression (10) is the absence of dependence on the waveguide depth. This circumstance is related to the fact that in the case under the consideration the amplitude of the exciting field at the output is determined by the expression:

$$E_{0,ZMW} = E_0 e^{-z/L} \quad (11)$$

where $L$ characterizes the attenuation of the TE$_{11}$ mode and z is measured from the excitation input point. For an atom located at point $z_A$, the radiation scattered by the atom can be described by the expression:

$$E_{sc} = C_{11}\alpha_A E_{in}(z_A)e^{-(z-H/2)/L} = C_{11}\alpha_A E_0 e^{-z_A/L}e^{-(z-z_A)/L} = C_{11}\alpha_A E_0 e^{-z/L} \quad (12)$$

where $C_{11}$ denotes the coefficient at the TE$_{11}$ mode in the mode expansion of the Green's function.

At the waveguide output ($z=H$) (11) and (12) have the same dependence on $H$ and therefore the effect of extraordinary transmission in a sufficiently deep nanowell should not depend on the thickness $H$.

More precise estimation of $\bar{\xi}$ in (9) for arbitrary ZMW depths can be found by numerical calculations. To quantify the effect of EOT, it is necessary to solve the complete system of Maxwell's equations with self-consistent consideration of the field scattered by an atom:

$$\begin{aligned} \nabla\times\nabla\times\mathbf{E}(\mathbf{r}) - k^2\varepsilon(\mathbf{r})\mathbf{E}(\mathbf{r}) &= J_{0,ext}(\mathbf{r}) + 4\pi k^2 \mathbf{P}_A \delta(\mathbf{r}-\mathbf{r}_A) \\ \mathbf{P}_A &= \ddot{\alpha}_A \mathbf{E}(\mathbf{r}_A) \end{aligned} \quad (13)$$

where $\mathbf{J}_{ext}(\mathbf{r})$ denotes the current that generates the external field $\mathbf{E}_{0,ZMW}(\mathbf{r})$, $\mathbf{P}_A$ denotes the induced dipole moment of the atom, $\varepsilon(\mathbf{r})$ denotes the permittivity of ZMW. Delta function $\delta(\mathbf{r})$ in the system of Eqs.(13) is difficult to implement in numerical calculations, since the size of the atom ($\sim r_{Bohr}$) is several orders of magnitude smaller than the characteristic dimensions of ZMW ($D_{ZMW}$, $H_{ZMW}$), which in turn is much smaller than the dimensions of the system with the detector ($L_{ext}$), $r_{Bohr} \ll D_{ZMW}, H_{ZMW} \ll L_{ext}$. Therefore, to solve (13) and estimate the geometric parameter $\bar{\xi}$, which does not depend on the characteristics of the atomic transition, we will simulate the oscillations of electrons in an atom by oscillations of electrons in a plasmonic meta-atom of radius $R_A$. From an electromagnetic point of view, the oscillations of electrons in a plasmonic meta-atom are completely identical to the oscillations of electrons in an atom. The polarizability of a plasmonic meta-atom has the form [5]

$$\alpha_{MA} = R_A^3 \frac{\varepsilon_{MA}-1}{\varepsilon_{MA}+2+2i(kR_A)^3}, kR_A \ll 1 \quad (14)$$

where $\varepsilon_{MA}$ denotes the permittivity of the meta-atom. The imaginary term in the denominator of (14) is related to radiation losses. It is easy to see that under Drude dispersion law $\varepsilon_{MA} = 1 - \omega_{pl}^2/\omega^2$ the polarizability of the plasmonic meta-atom takes the form:

$$\alpha_{MA} = \frac{3}{4k^3}\frac{\Gamma}{\omega-\omega_0+i\Gamma/2}; \omega_0 = \frac{\omega_{pl}}{\sqrt{3}}; \Gamma = \omega_{pl}\frac{2(kR_A)^3}{3\sqrt{3}} \qquad (15)$$

that is, it coincides with Eq.(2) with the only difference that the resonant frequency and decay rate are determined by the plasmon frequency $\omega_{pl}$ and radius $R_A$. Since a meta-atom is an auxiliary object, it is easy to select parameters that provide an approximation of Eq. (2) for any specific atom or other quantum object with low internal losses. Additional tuning of the meta-atom parameters is also possible by using a dispersion law that is more complex than the Drude law.

As an example, below we will consider a meta-atom with a radius $R_A$=3nm and a resonant wavelength of 532 nm. In this case, the approximation (14),(15) is very good, although it must be borne in mind that the resonance can shift compared to free space due to reflections from the walls of ZMW (see Eq.(3)).

The results of simulations within the Comsol Multiphysics [32] are shown in Figs.2-4 for ZMWs in an Al film of thickness $H_{ZMW}$= 100 nm, freely standing in the vacuum (see Fig. 1). The permittivity of Al is taken from [33].

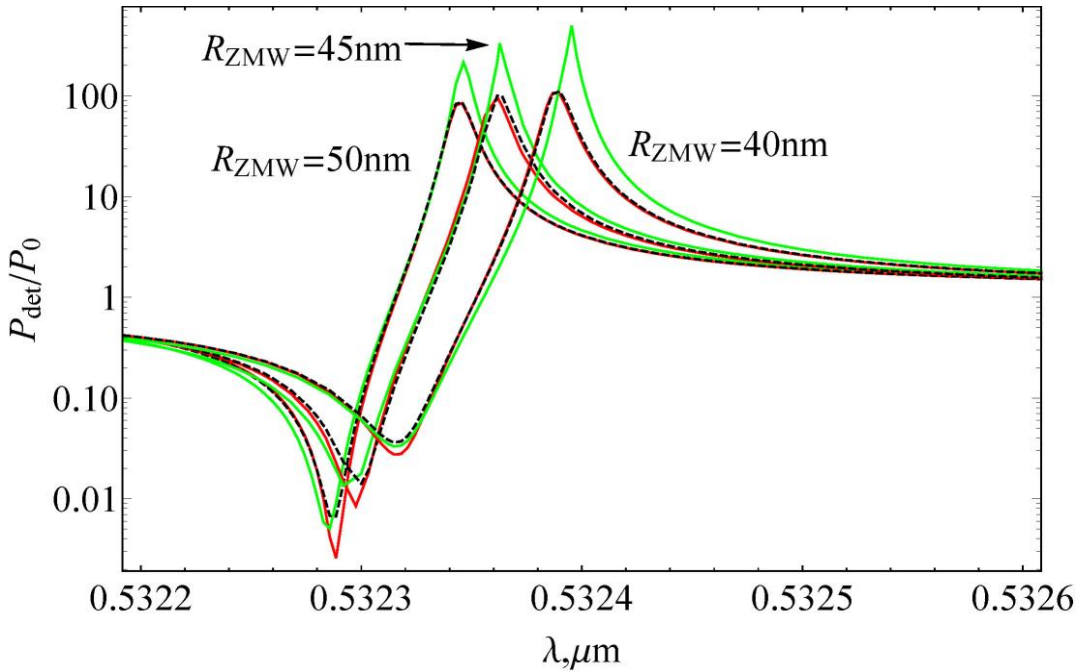

Fig.2. Light transmission coefficient through ZMWs with radii $R_{ZMW}$=40nm, 45nm, 50nm. The atom is located near the bottom of ZMW ($Z_A$=6 nm, red), in the middle of ZMW ($Z_A$= 50 nm, green), and at the top of ZMW ($Z_A$= 94 nm, black dashed curve).

From Fig.2 it is clear that in this system the effect of EOT does occur and that the maximum of EOT weakly depends on the ZMW radius. From these figures it is also clear that the signal in the detector has the form of a classical Fano resonance [34], which is not surprising since there is an interaction between a narrow resonance in the atom and a very broad resonance in the ZMW (see Eq.(9)).

Just below resonance ($\lambda \approx 0.53230$ μm) for any position of the atom in ZMW, there is a significant blocking of transmission (by 2 orders of magnitude or more).

With a different frequency detuning (slightly higher than resonance, $\lambda \approx 0.53235$ μm) the situation is reversed, since the phase of the field scattered by the atom changes by π. In this case, EOT occurs, the maximum of which weakly depends on the position of the atom inside the ZMW and on the radius of the ZMW. The shapes of the transmission spectra in Fig.2 are well described by the Eq.(9) with $\bar{\xi} \approx 2$. It is also important that in Fig.2, despite the anisotropy of the system, for each size of a meta-atom there is only 1 resonance. This happens because axial symmetry is preserved on the axis of the system and splitting is possible only for the z orientation of the transition dipole moment, which is not excited, since in our case there is no z component of the external field on the axis. Note that if we normalize the detected power to the incident flux, considering reflection from the metal film, rather than to the flux in the detector without an atom, then the transmission coefficient will be about 60%.

The influence of ZMW depth on the extraordinary transmission effect is more significant (see Fig. 3).

It is evident from Fig. 3 that with increasing ZMW thickness the maximum transmittance increases, but this increase is saturated for $H_{ZMW} \to \infty$, which is due to the fact that in this case the effect is described by one, least damped mode (see (10), (11), (12)). For ZMW with perfectly conducting walls, the transmittance becomes significantly higher (black dashed line in Fig. 3). Fig.4 shows the dependence of the maximum transmittance on thickness, where saturation is especially clearly visible.

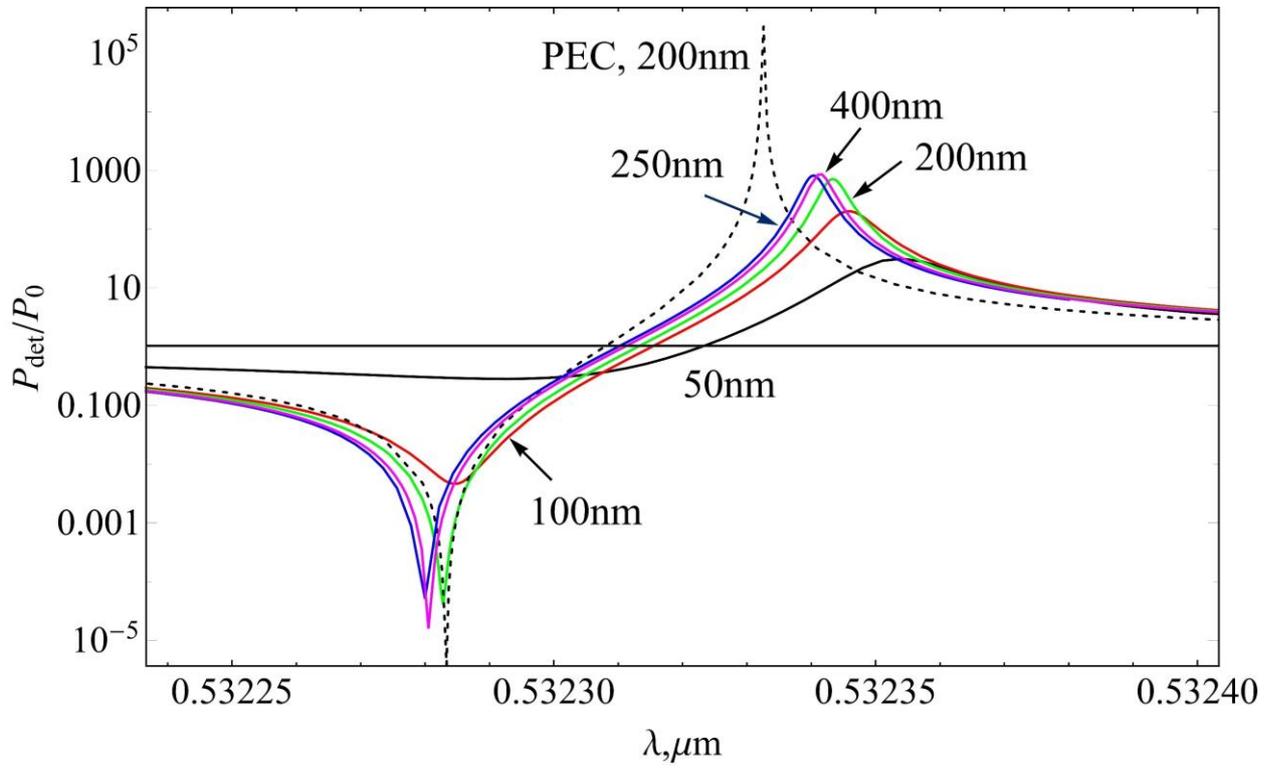

Fig. 3. Transmission coefficient as a function of wavelength for different ZMW thicknesses. The black dashed line corresponds to ZMW with perfectly conducting walls. In all cases, the atom is located in the middle of the ZMW with a diameter of $D_{ZMW}$=100 nm.

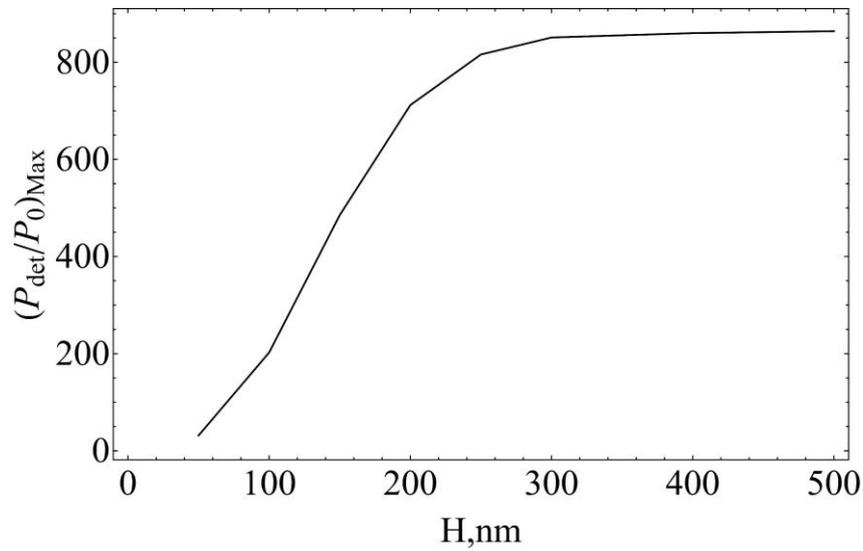

Fig. 4. Dependence of the maximum transmission on the ZMW thickness. $D_{ZMW}$=100 nm;

To clarify the nature of the effects of extraordinary transmission and blocking of light transmission through ZMW, Fig. 5 shows the Poynting vector streamlines.

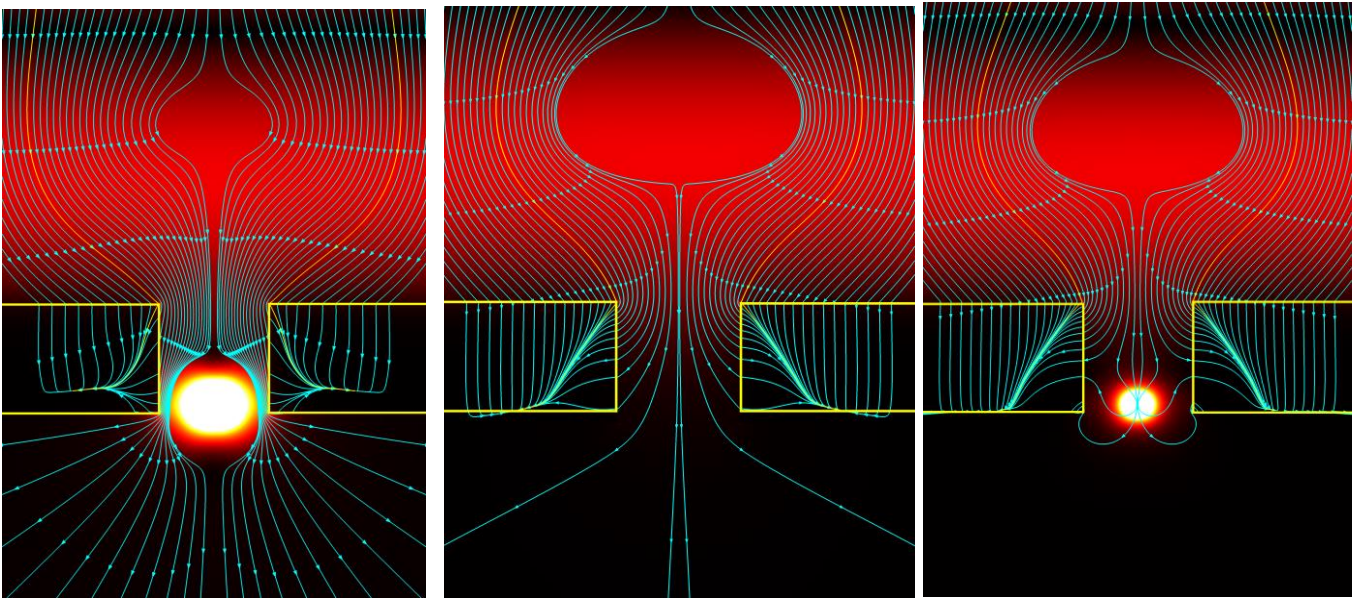

Fig.5. Streamlines of the Poynting vector in case: (a) λ=0.53235 μm; (b) λ=0.53229 μm (no atom) and (c) λ =0.53229 μm ($Z_A$=6nm, $D_{ZMW}$=100nm).

From Fig. 5 it is clear that, judging by the streamlines arriving at the edge of the hole (yellow curves), in the case shown in Fig. 5(a), the area from which the exciting field flow is collected (~$\lambda^2$) is larger in area than in the case of the absence of an atom (Fig. 5(b)), which in turn is greater than the energy flux captured in the case shown in Fig. 5(c).

Both in the absence of an atom (Fig. 5(b)), and in the case λ=0.53229 μm (Fig. 5(c)), light propagation in ZMW leads to almost complete absorption in the walls. Moreover, Fig. 5(c) shows that at the bottom of the ZMW, power enters the ZMW instead of the intuitive view that light should exit from the ZMW. Therefore, determining the transmission coefficient based on the power transmitted directly through the bottom of the waveguide, that is, in the near field of ZMW, cannot be used as the basis for experimental measurements. The geometry shown in Fig. 1 is designed for far-field measurements and therefore correctly describes the transmission of light.

In case λ=0.53235μm (Fig. 5(a)) there is almost no absorption in the walls, which leads to a further increase of EOT. Figuratively speaking, the atom draws energy into ZMW from an area of the order of $\lambda^2$ and almost completely transfers it to the detector.

From a physical point of view, the nature of the discovered effect may also lie in the fact that placing an atom with a large polarizability into a ZMW can be considered as a change in the topology of the ZMW and the transformation of the ZMW into a coaxial waveguide in which there is no cutoff frequency [36]-[39]. In turn, the absence of a cutoff frequency allows light to propagate without exponential decay, and thereby significantly increase the transmission of light through the ZMW. Note, however, that this argument is preliminary since a detailed analysis of the polarization structure of the fields inside the ZMW is still to come.

Above, we considered the case of an atom located on the axis, but the atom can also be located outside ZMW axis. In this case, the atom in ZMW ceases to be an isotropic quantum system and the three times degenerate oscillation is split into 3 different frequencies in the general case (see Eq.(3)).

The most noticeable splitting is for oscillations in a *x-y* plane perpendicular to the ZMW axis, that is, the oscillations along ($\rho$) and across ($\varphi$) the radius have different frequencies, which are qualitatively described far from the axis by the expressions [40]

$$\frac{\Delta\lambda}{\lambda_0} \sim \frac{3}{4}\text{Re}\left(\frac{\varepsilon-1}{\varepsilon+1}\right)\frac{R_{ZMW}^3}{(R_{ZMW}-\rho)^3}(\mathbf{d}\perp \text{ to wall})$$

$$\frac{\Delta\lambda}{\lambda_0} \sim \frac{3}{8}\text{Re}\left(\frac{\varepsilon-1}{\varepsilon+1}\right)\frac{R_{ZMW}^3}{(R_{ZMW}-\rho)^3}(\mathbf{d}\| \text{ to wall})$$

(16)

and therefore, maximum transmission will occur at lower frequencies than shown on Fig. 2 - Fig. 5.

Fig. 6 shows the spectra of light transmission through ZMW at various atomic distances from the *Z*-axis.

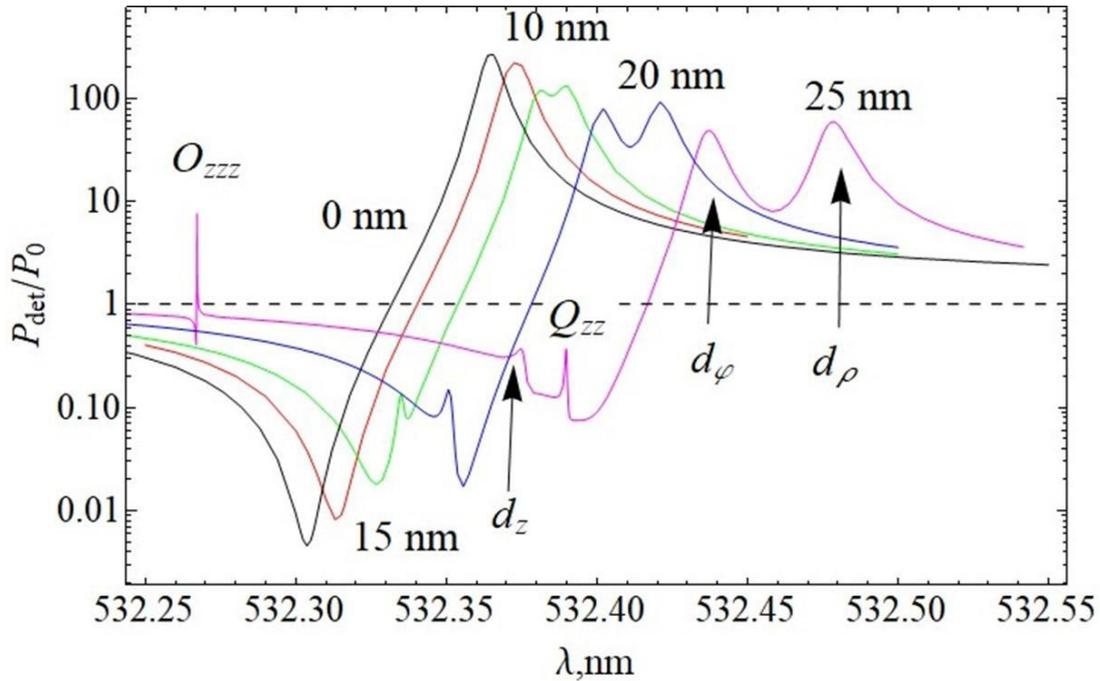

Fig.6. Transmission coefficient for various atomic displacements along the *X* axis ($X_A$= 0nm (atom on axis) - black, $X_A$=10 nm – red, $X_A$=15nm - green, $X_A$=20 nm - blue, $X_A$=25nm - magenta). $Z_A$=50 nm. $D_{ZMW}$=100nm.

From Fig. 6 it is clear that, indeed, breaking the symmetry of the system leads to splitting of dipole resonance lines and even to the excitation of multipole transitions of higher orders (quadrupoles and octupoles).

The left main maximum of EOT corresponds to φ orientation of the transition dipole moment, while the right one corresponds to ρ orientation of the transition dipole moment. Small maxima to the left of the main ones correspond to the excitation of the *z*-oriented electric dipole moment of the transition and higher-order multipoles. This splitting of resonances when the atom is displaced from the axis can be used to track the position of the atom.

In conclusion, a theoretical framework has been developed to describe the influence of resonant atoms on the transmission of light through ZMW. It is shown that the predicted effect of extraordinary light transmission has a non-trivial physical nature, which is caused by a strong interaction of evanescent fields in ZMW and the atom. As a result of this interaction, both Fano resonances and splitting of extraordinary transmission lines arise in the transmission spectra due to the anisotropy of the system. The results obtained will make it possible not only to determine the presence of an atom in the ZMW, but also its location.

An important result of the work is that the scattering cross-section of an atom in ZMW ($\sim\lambda^2$) coincides in order of magnitude with the maximum resonant cross-section of an atom in free space ($\sim\lambda^2$).

The most clearly predicted effect can be observed for an atom at rest, that is, for an atom that is in the ZMW for a time $\tau$ that is greater than the time $\tau_0$ of spontaneous decay $\tau = H_{ZMW} / V_{Atom} > \tau_0$. This condition is met, for example, for cold Rb atoms moving at a speed of $V_{Atom} \sim 1$ m/s. De Broglie wavelength $\lambda_{dB}$ of such atoms is also quite small compared to the size of ZMW, $\lambda_{dB} \sim 5$nm $\ll D_{ZMW}, H_{ZMW}$, which allows us to consider their dynamics as classical.

To demonstrate the effect of EOT, one can also consider quantum systems in crystalline matrices (NV centers in diamond, etc.). In the latter case, the found effect can be used to switch the transmission of light with a small change in wavelength or geometry of the system.

The predicted effects of extraordinary light transmission and blocking are of a completely general nature and thus open new possibilities for studying the interaction of light with quantum objects in nanoenvironment and the development of new quantum technologies.

**Acknowledgments**—The author acknowledges financial support from the Russian Science Foundation (Grant No. 23-42-00049).

*klimov256@gmail.com